\documentclass[12pt]{article}

\usepackage{amsmath}
\usepackage{amssymb}
\usepackage{amsfonts}
\usepackage{latexsym}
\usepackage{color}

\catcode `\@=11 \@addtoreset{equation}{section}

\catcode `\@=12

\newcommand{\be}{\begin{equation}}
\newcommand{\en}{\end{equation}}
\newcommand{\bea}{\begin{eqnarray}}
\newcommand{\ena}{\end{eqnarray}}
\newcommand{\beano}{\begin{eqnarray*}}
\newcommand{\enano}{\end{eqnarray*}}
\newcommand{\bee}{\begin{enumerate}}
\newcommand{\ene}{\end{enumerate}}

\newcommand{\N}{\mathfrak N}

\newcommand{\mc}{\mathcal}

\newcommand{\D}{{\mc D}}

\newcommand{\F}{{\cal F}}

\newcommand{\Lc}{{\cal L}}

\newcommand{\1}{1 \!\! 1}

\newcommand{\Hil}{\mc H}

\newtheorem{thm}{Theorem}

\catcode `\@=11 \@addtoreset{equation}{section}
\catcode `\@=12

\begin{document}

\thispagestyle{empty}

\vspace*{2cm}

\begin{center}
{\Large \bf  Modified Landau levels, damped harmonic oscillator and two-dimensional pseudo-bosons}\\[10mm]

{\large S. Twareque Ali} \footnote[1]{Department of Mathematics
and Statistics, Concordia University,
Montr\'eal, Qu\'ebec, CANADA H3G 1M8\\
e-mail: stali@mathstat.concordia.ca}
\vspace{3mm}\\

{\large F. Bagarello} \footnote[2]{ Dipartimento di Metodi e
Modelli Matematici,
Facolt\`a di Ingegneria, Universit\`a di Palermo, I-90128  Palermo, ITALY\\
e-mail: bagarell@unipa.it\,\,\,\, Home page:
www.unipa.it/$^\sim$bagarell}
\vspace{3mm}\\

{\large Jean Pierre Gazeau}\footnote[3]{Laboratoire APC,
Universit\'e Paris 7-Denis Diderot, 10, rue A. Domon et L. Duquet,
75205 Paris Cedex 13, France}
\end{center}

\vspace*{2cm}

\begin{abstract}
\noindent In a series of recent papers one of us has analyzed in
some details a class of elementary excitations called {\em
pseudo-bosons}. They arise from a special deformation of the
canonical commutation relation $[a,a^\dagger]=\1$, which is
replaced by $[a,b]=\1$, with $b$ not necessarily equal to $a^\dagger$.
Here, after a
two-dimensional extension of the general framework, we apply the
theory to a generalized version of the two-dimensional Hamiltonian
describing Landau levels. Moreover, for this system, we discuss
coherent states and we deduce a resolution of the identity. We
also consider a different class of examples arising from a classical system,
i.e. a damped harmonic oscillator.

\end{abstract}

\vspace{2cm}

%{\bf PACS Numbers}:  .......

\vfill

%\pagenumbering{roman}

\newpage

\section{Introduction}

In a series of recent papers \cite{bagpb1,bagpb2,bagpb3,bagpb4}, one of
us (FB) has investigated some mathematical aspects of the
so-called {\em pseudo-bosons}, originally introduced by Trifonov
in \cite{tri}. They arise from the canonical commutation relation
$[a,a^\dagger]=\1$ upon replacing $a^\dagger$ by another (unbounded)
operator $b$ not (in general) related to $a$: $[a,b]=\1$. We have
shown that $N=ba$ and $N^\dagger=a^\dagger b^\dagger$ can be both
diagonalized, and that their spectra coincide with the set of
natural numbers (including 0), ${\Bbb N}_0$. However the sets of
related eigenvectors are not orthonormal (o.n) bases but,
nevertheless, they are automatically {\em biorthogonal\/}. In all the
examples considered so far, they are bases of the Hilbert space of the system,
$\Hil$, and, in some cases, they turn out to be {\em Riesz bases\/}.

To our knowledge, not many physical consequences of
this construction have been
discussed up to now. For this reason, extending what two of us
(STA and FB) have already done in \cite{alibag}, we will construct
here a two-dimensional model which fits the main assumptions of
the construction given in \cite{bagpb1} and which is physically
motivated. We will further consider a second example, again
physically motivated, arising from the quantization of the damped
harmonic oscillator, \cite{ban}.

This paper is organized as follows:  in the next section we
introduce and discuss two-dimensional pseudo-bosons analyzing some
of their mathematical properties and their related coherent
states.  In Section III we introduce the {\em generalized Landau
levels} (GLL) and we discuss them in the context of Section II.
Section IV is devoted to our analysis of the quantum damped harmonic
oscillator, while Section V contains our conclusions.

\section{The commutation rules}

In this section we will construct a two-dimensional (2-D) version  of
what originally proposed in \cite{bagpb1}, to which we refer for
further comments on the 1-D situation.

Let $\Hil$ be a given Hilbert space with scalar product
$\left<.,.\right>$ and related norm $\|.\|$. We introduce two
pairs of operators, $a_j$ and $b_j$, $j=1,2$, acting on $\Hil$ and
satisfying the following commutation rules \be [a_j,b_j]=\1, \quad
\mbox{ and }\quad
[a_1^\sharp,a_2^\sharp]=[a_1^\sharp,b_2^\sharp]=[b_1^\sharp,b_2^\sharp]=0,
\label{21} \en where $x^\sharp$ stands for $x$ or $x^\dagger$
($x=a_j, b_j$). Of course, they collapse to the CCR's for
independent modes if $b_j=a^\dagger_j$, $j=1,2$. It is well known
that $a_j$ and $b_j$ are unbounded operators, so they cannot be
defined on all of $\Hil$. Following \cite{bagpb1}, and writing
$D^\infty(X):=\cap_{p\geq0}D(X^p)$ (the common  domain of all the powers of the
operator $X$), we consider the
following:

\vspace{2mm}

{\bf Assumption 1.--} there exists a non-zero
$\varphi_{0,0}\in\Hil$ such that $a_j\varphi_{0,0}=0$, $j=1,2$,
and $\varphi_{0,0}\in D^\infty(b_1)\cap D^\infty(b_2)$.

{\bf Assumption 2.--} there exists a non-zero $\Psi_{0,0}\in\Hil$
such that $b_j^\dagger\Psi_{0,0}=0$, $j=1,2$, and $\Psi_{0,0}\in
D^\infty(a_1^\dagger)\cap D^\infty(a_2^\dagger)$.

\vspace{2mm}

Under these assumptions we can introduce the following vectors in
$\Hil$: \be
\varphi_{n,l}=\frac{1}{\sqrt{n!l!}}\,b_1^n\,b_2^l\,\varphi_{0,0}
\quad \mbox{ and }\quad
\Psi_{n,l}=\frac{1}{\sqrt{n!l!}}(a_1^\dagger)^n(a_2^\dagger)^l\Psi_{0,0},
\quad n,l\geq 0. \label{22}\en Let us now define the unbounded
operators $N_j:=b_ja_j$ and $\N_j:=N_j^\dagger=a_j^\dagger
b_j^\dagger$, $j=1,2$.  It is possible to check that
$\varphi_{n,l}$ belongs to the domain of $N_j$, $D(N_j)$, and
$\Psi_{n,l}\in D(\N_j)$, for all $n,l\geq0$ and $j=1,2$. Moreover,
\be N_1\varphi_{n,l}=n\varphi_{n,l}, \quad
N_2\varphi_{n,l}=l\varphi_{n,l}, \quad \N_1\Psi_{n,l}=n\Psi_{n,l},
\quad \N_2\Psi_{n,l}=l\Psi_{n,l}. \label{23}\en

Under the above assumptions it is  easy to check that
$\left<\Psi_{n,l},\varphi_{m,k}\right>=\delta_{n,m}\delta_{l,k}\left<\Psi_{0,0},\varphi_{0,0}\right>$
for all $n, m, l, k\geq0$, which, if we chose the normalization of
$\Psi_{0,0}$ and $\varphi_{0,0}$ in such a way that
$\left<\Psi_{0,0},\varphi_{0,0}\right>=1$, becomes \be
\left<\Psi_{n,l},\varphi_{m,k}\right>=\delta_{n,m}\delta_{l,k},
\quad \forall n,m,l,k\geq0. \label{27}\en This means that the sets
$\F_\Psi=\{\Psi_{n,l},\,n,l\geq0\}$ and
$\F_\varphi=\{\varphi_{n,l},\,n,l\geq0\}$ are {\em biorthogonal} and,
because of this, the vectors of each set are linearly independent.
If we now call $\D_\varphi$ and $\D_\Psi$ respectively the linear
span of  $\F_\varphi$ and $\F_\Psi$, and $\Hil_\varphi$ and
$\Hil_\Psi$ their closures, then \be f=\sum_{n,l=0}^\infty
\left<\Psi_{n,l},f\right>\,\varphi_{n,l}, \quad \forall
f\in\Hil_\varphi,\qquad  h=\sum_{n,l=0}^\infty
\left<\varphi_{n,l},h\right>\,\Psi_{n,l}, \quad \forall
h\in\Hil_\Psi. \label{210}\en What is not in general ensured is
that the Hilbert spaces introduced so far all coincide, i.e. that
$\Hil_\varphi=\Hil_\Psi=\Hil$. Indeed, we can only state that
$\Hil_\varphi\subseteq\Hil$ and $\Hil_\Psi\subseteq\Hil$. However,
motivated by the examples already discussed in the literature and
anticipating the discussion in Section III,  we make the

\vspace{2mm}

{\bf Assumption 3.--} The above Hilbert spaces all coincide:
$\Hil_\varphi=\Hil_\Psi=\Hil$,

\vspace{2mm}

which was introduced in \cite{bagpb1}. This means, in particular,
that both $\F_\varphi$ and $\F_\Psi$ are bases of $\Hil$. Let us
now introduce the operators $S_\varphi$ and $S_\Psi$ via their
action respectively on  $\F_\Psi$ and $\F_\varphi$: \be
S_\varphi\Psi_{n,k}=\varphi_{n,k},\qquad
S_\Psi\varphi_{n,k}=\Psi_{n,k}, \label{213}\en for all $n,
k\geq0$, which also imply that
$\Psi_{n,k}=(S_\Psi\,S_\varphi)\Psi_{n,k}$ and
$\varphi_{n,k}=(S_\varphi \,S_\Psi)\varphi_{n,k}$, for all
$n,k\geq0$. Hence \be S_\Psi\,S_\varphi=S_\varphi\,S_\Psi=\1 \quad
\Rightarrow \quad S_\Psi=S_\varphi^{-1}. \label{214}\en In other
words, both $S_\Psi$ and $S_\varphi$ are invertible and one is the
inverse of the other. Furthermore, we can also check that they are
both positive, well defined and symmetric, \cite{bagpb1}. Moreover, at
least formally, it is possible to write these operators in the
bra-ket notation as \be S_\varphi=\sum_{n,k=0}^\infty
|\varphi_{n,k}><\varphi_{n,k}|,\qquad S_\Psi=\sum_{n,k=0}^\infty
|\Psi_{n,k}><\Psi_{n,k}|. \label{212}\en
 These expressions are
only formal, at this stage, since the series may not converge in
the uniform topology and the operators $S_\varphi$ and $S_\Psi$ could be unbounded.
This aspect was exhaustively discussed in \cite{bagpb1}, where the role
of Riesz bases\footnote{Recall that a set of vectors $\phi_1, \phi_2 , \phi_3 , \; \ldots \; ,$ is a Riesz basis of a Hilbert space $\mathcal H$, if there exists a bounded operator $V$, with bounded inverse, on $\mathcal H$, and an orthonormal basis of $\Hil$,  $\varphi_1, \varphi_2 , \varphi_3 , \; \ldots \; ,$ such that $\phi_j=V\varphi_j$, for all $j=1, 2, 3,\ldots$} in relation with the boundedness of $S_\varphi$ and
$S_\Psi$ has been discussed in detail. We shall come back to this
aspect later. However, we shall not assume here, except when
explicitly stated, what has been called {\em Assumption 4} in
\cite{bagpb1}, since in most examples considered so far, and in
what we are going to discuss in Section III, this assumption is not satisfied.

It is interesting to remark that, as in \cite{bagpb1}, even these
two-dimensional pseudo-bosons give rise to interesting
intertwining relations among non self-adjoint operators, see
\cite{bagpb3} and references therein. In particular it is easy to
check that \be S_\Psi\,N_j=\N_jS_\Psi \quad \mbox{ and }\quad
N_j\,S_\varphi=S_\varphi\,\N_j, \label{219}\en $j=1,2$. This is
related to the fact that the spectra of, say, $N_1$ and $\N_1$
coincide and that their eigenvectors are related by the operators
$S_\varphi$ and $S_\Psi$, in agreement with the literature on
intertwining operators, \cite{intop,bag1}, and on pseudo-Hermitian
quantum mechanics, see \cite{mosta} and references therein.

\subsection{Coherent states}

As it is well known there exist several different, and not always
equivalent, ways to define {\em
coherent states}, \cite{book1,book2}. In this paper we will
adopt the following definition,  generalizing
\cite{bagpb1}. Let $z_1$ and $z_2$ be two complex variables, $z_1,
z_2\in \D$ (some domain in $\Bbb{C}$), and let us introduce the following
operators: \be
U_j(z_j)=e^{z_j\,b_j-\overline{z}_j\,a_j}=e^{-|z_j|^2/2}\,e^{z_j\,b_j}\,e^{-\overline{z}_j\,a_j},
\quad
V_j(z_j)=e^{z_j\,a_j^\dagger-\overline{z}_j\,b_j^\dagger}=e^{-|z_j|^2/2}\,e^{z_j\,a_j^\dagger}\,e^{-\overline{z}_j\,b_j^\dagger},
\label{31}\en $j=1,2$, and \be
U(z_1,z_2):=U_1(z_1)\,U_2(z_2),\qquad
V(z_1,z_2):=V_1(z_1)\,V_2(z_2), \label{31b}\en and the following
vectors: \be \varphi(z_1,z_2)=U(z_1,z_2)\varphi_{0,0},\qquad
\Psi(z_1,z_2)=V(z_1,z_2)\,\Psi_{0,0}. \label{32}\en \vspace{2mm}

{\bf Remarks:--} (1) Due to the commutation rules for the
operators $b_j$ and $a_j$, we  clearly have
$[U_1(z_1),U_2(z_2)]=[V_1(z_1),V_2(z_2)]=0$.

(2) Since the operators $U$ and $V$ are, for generic $z_1$ and
$z_2$, unbounded, definition (\ref{32}) makes sense only if
$\varphi_{0,0}\in D(U)$ and $\Psi_{0,0}\in D(V)$, a condition which
will be assumed here. In \cite{bagpb1} it was proven that, for
instance, this is so when $\F_\varphi$ and $\F_\Psi$ are Riesz
bases.

(3) The set $\D$ could,  in
principle, be a proper subset of $\Bbb{C}$.

\vspace{2mm}

It is possible to write the vectors $\varphi(z_1,z_2)$ and
$\Psi(z_1,z_2)$ in terms of the vectors of $\F_\Psi$ and
$\F_\varphi$ as \be
\varphi(z_1,z_2)=e^{-(|z_1|^2+|z_2|^2)/2}\,\sum_{n,l=0}^\infty\,\frac{z_1^nz_2^l}{\sqrt{n!\,l!}}\,\varphi_{n,l},
\quad
\Psi(z_1,z_2)=e^{-(|z_1|^2+|z_2|^2)/2}\,\sum_{n,l=0}^\infty\,\frac{z_1^nz_2^l}{\sqrt{n!\,l!}}\,\Psi_{n,k}.
\label{33}\en

These vectors are called {\em coherent} since they are eigenstates
of the lowering operators. Indeed we can check that \be
a_j\varphi(z_1,z_2)=z_j\varphi(z_1,z_2), \qquad
b_j^\dagger\Psi(z_1,z_2)=z_j\Psi(z_1,z_2), \label{34}\en for
$j=1,2$ and $z_j\in\D$. It is also a standard exercise, putting
$z_j=r_j\,e^{i\theta_j}$, to check that the following operator
equalities hold: \be
\frac{1}{\pi^2}\int_{\Bbb{C}}\,dz_1\int_{\Bbb{C}}\,dz_2\,
|\varphi(z_1,z_2)><\varphi(z_1,z_2)|=S_\varphi, \quad
\frac{1}{\pi^2}\int_{\Bbb{C}}\,dz_1\int_{\Bbb{C}}\,dz_2\,
|\Psi(z_1,z_2)><\Psi(z_1,z_2)|=S_\Psi, \label{35}\en as well as
\be \frac{1}{\pi^2}\int_{\Bbb{C}}\,dz_1\int_{\Bbb{C}}\,dz_2\,
|\varphi(z_1,z_2)><\Psi(z_1,z_2)|=
\frac{1}{\pi^2}\int_{\Bbb{C}}\,dz_1\int_{\Bbb{C}}\,dz_2\,
|\Psi(z_1,z_2)><\varphi(z_1,z_2)|=\1, \label{36}\en which are
written in convenient bra-ket notation. It should be said that
these equalities are, most of the times, only formal results.
Indeed it is not difficult to construct examples in which {\em
something goes wrong} and, for instance, the resolution of the
identity for the pair $\varphi(z_1,z_2)$ and $\Psi(z_1,z_2)$ does not hold as expected. As the
following theorem will show, this is a reflection of the fact that
the operators $S_\varphi$ and $S_\Psi$ are unbounded, or,
equivalently, of the fact that $\F_\varphi$ and $\F_\Psi$ are not
Riesz bases. Indeed we have the following general result, which
was essentially stated in \cite{bagpb2} for a concrete example of
1-D pseudo-bosons, and which we extend here to the general
setting.

\vspace{2mm}

\begin{thm} Let $a_j$, $b_j$, $\F_\varphi$, $\F_\Psi$, $\varphi(z_1,z_2)$ and $\Psi(z_1,z_2)$ be as above. Let us assume that
(1) $\F_\varphi$, $\F_\Psi$ are Riesz bases; (2) $\F_\varphi$,
$\F_\Psi$ are biorthogonal. Then (\ref{36}) holds true.

\end{thm}

The proof of this theorem does not differ significantly  from that
given in \cite{bagpb2}, so that it will not be repeated here. The
meaning of the theorem is the following: suppose that following
the above construction the coherent states we get do not produce a
resolution of the identity. Then, since $\F_\varphi$ and
$\F_\Psi$ are automatically biorthogonal, they cannot be Riesz
bases (neither one of them)! However, this theorem does not hold in
general for other types of coherent states. We will come back on this point in the next
section.

\section{Generalized Landau levels}

The Hamiltonian of a single electron, moving on a two-dimensional
plane and subject to a uniform magnetic field along the
$z$-direction, is given by the operator
\begin{equation}
H_0'={\frac 12}\,\left(\underline p+\underline A(r)\right)^2={\frac
12}\; \left(p_x-{\frac y2}\right)^2+{\frac 12}\,\left(p_y+{\frac
x2} \right)^2, \label{41}
\end{equation}
where we have used minimal coupling and the symmetric gauge $\vec
A=\frac{1}{2}(-y,x,0)$.

The spectrum of this Hamiltonian is easily obtained by first
introducing the new variables
  \be
\label{42}
  P_0'= p_x-y/2, \hspace{5mm}     Q_0'= p_y+x/2.
  \en
In terms of $P_0'$ and $Q_0'$ the single electron hamiltonian,
$H_0$, can be rewritten as
 \be
\label{43}
  H_0'=\frac{1}{2}(Q_0'^2 + P_0'^2).
  \en
On a classical level, the transformation (\ref{42}) is part of a canonical map from the
phase space variables $(x,y,p_x,p_y)$ to $(Q_0,P_0,Q_0',P_0')$,
where
 \be
\label{44}
   P_0= p_y-x/2, \hspace{5mm}
  Q_0= p_x+y/2,
   \en
which can be used to construct a second hamiltonian $ H_0=\frac{1}{2}(Q_0^2 + P_0^2).$

  The corresponding quantized  operators  satisfy the commutation relations:
$$ [x, p_x] = [y, p_y] = i, \quad [x,p_y] = [y,p_x] = [x,y] = [p_x , p_y ] = 0, $$
and
  \be
\label{45}
 [Q_0,P_0] = [Q_0',P_0']=i, \quad  [Q_0,P_0']=[Q_0',P_0]=[Q_0,Q_0']=[P_0,P_0']=0,
  \en
so that $[H_0,H_0']=0$.

We refer to \cite{b6} and references therein for a discussion on how the corresponding wave functions look in different representations.   In \cite{alibag} two of us
(STA and FB) have considered, in the context of supersymmetric
(SUSY) quantum mechanics, an extended version of $H_0'$, an extension
needed due to the fact that for the Hamiltonian of the standard Landau levels
(SLL) there is essentially no difference between $H_0'$ and its
SUSY partner Hamiltonian.

The extension constructed in \cite{alibag} is very natural and
simple:  introducing the vector valued function $\vec
W_0=-\frac{1}{2}(x,y,0)=(W_{0,1},W_{0,2},0)$, we may rewrite the
operators in (\ref{42}) and (\ref{44}) as \be
P_0'=p_x+W_{0,2},\hspace{4mm}Q_0'=p_y-W_{0,1},\hspace{4mm}
P_0=p_y+W_{0,1},\hspace{4mm}Q_0=p_x-W_{0,2}.\label{48}\en This
definition was extended in \cite{alibag} as follows: \be
p'=p_x+W_{2},\hspace{4mm}q'=p_y-W_{1},\hspace{4mm}
p=p_y+W_{1},\hspace{4mm}q=p_x-W_{2},\label{49}\en introducing a
vector superpotential $\vec W=(W_{1},W_{2},0)$.

\vspace{1mm}

Here, since we are interested in constructing 2-D pseudo-bosons, it is
convenient to introduce two (in general) complex and different vector
superpotentials (this is a slight abuse of language!) $\vec
W=(W_1,W_2)$ and $\vec V=(V_1,V_2)$, and we put \be
P'=p_x+W_{2},\hspace{4mm}Q'=p_y-W_{1},\hspace{4mm}
P=p_y+V_{1},\hspace{4mm}Q=p_x-V_{2}.\label{410}\en Our notation is
the following: all operators with suffix $0$ are related to
the SLL. The same operators, without the $0$, have to do with our
generalized model, i.e. with the GLL. Notice that these operators are, in general, not self-adjoint. Hence, while for example $P_0=P_0^\dagger$, we may have $P\neq P^\dagger$, depending on the choice of $V_1$.  The superpotentials should
 also be chosen in such a way that, first of all, $Q$, $P$, $Q'$ and
$P'$ satisfy the same commutation rules (\ref{45}) as their
$0$-counterparts:
  \be
\label{412}
 [Q,P] = [Q',P']=i, \quad  [Q,P']=[Q',P]=[Q,Q']=[P,P']=0.
  \en
These impose certain conditions on $\vec V$ and $\vec W$: \be
W_{1,x}=V_{2,y},\quad W_{2,x}=-V_{2,x},\quad
W_{1,y}=-V_{1,y},\quad W_{2,y}=V_{1,x}, \label{413}\en
as well as
\be V_{1,x}+V_{2,y}=W_{1,x}+W_{2,y}=-1. \label{414}\en
The subscripts $x,y$ denote differentiation with
respect to that variable.
Hence, as
it was already clear at the beginning, the two different vector
superpotentials must be related to each other. Notice that the
standard choice trivially satisfies all these conditions. We now
introduce the following operators: \be A'=\alpha'(Q'+i\,P'),\quad
B'=\gamma'(Q'-i\,P'),\quad A=\alpha (Q+i\, P),\quad B=\gamma (Q -i
P), \label{415}\en where $\alpha\,\gamma=\frac{1}{2}$ and
$\alpha'\,\gamma'=\frac{1}{2}$. Incidentally, we recall that for the
SLL the same linear combinations as in (\ref{415}) hold
with $\alpha=\alpha'=\gamma=\gamma'=\frac{1}{\sqrt{2}}$ and with
the operators $Q, P, Q'$ and $P'$ replaced respectively by $Q_0,
P_0, Q_0'$ and $P_0'$. Thus, the operators generalizing the Landau
Hamiltonians in \cite{alibag} are \be h'={\frac 12}\;
\left(p_x+W_2\right)^2+{\frac 12}\,\left(p_y-W_1\right)^2, \qquad
h={\frac 12}\; \left(p_x-V_2\right)^2+{\frac
12}\,\left(p_y+V_1\right)^2, \label{415b}\en which can be
rewritten as \be h'=B'A'-\frac{1}{2}\,\1,\qquad
h=BA-\frac{1}{2}\1. \label{416}\en The operators in (\ref{415})
are pseudo-bosonic since they satisfy the following commutation
rules: \be [A,B]=[A',B']=\1,\label{417}\en while all the other
commutators are trivial. It is important to observe that, since
$A^\dagger=\overline{\alpha}(Q^\dagger-iP^\dagger)$, and since $Q$
and $P$ are not necessarily self-adjoint, in general $B\neq
A^\dagger$. Analogously, in general $B'\neq {A'}^\dagger$.
Similar conclusions can be deduced starting from the pairs  $B^\dagger$, $A^\dagger$ and $B'^\dagger$, $A'^\dagger$.

At this stage it is interesting to say few words on the physical
meaning of our model. In other words: what is the physical
meaning of going from the SLL to these GLL? The answer is the following: suppose we interpret $\vec W$ and $\vec V$ in (\ref{415b}) as two different (but related) vector potentials describing two possibly different magnetic fields. These potentials are
$\vec A_\uparrow=(W_2,-W_1,0)$ for $h'$ and $\vec A_\downarrow=(-V_2,V_1,0)$ for $h$ (the reason
for this notation will be clear in a moment). Now, computing the associated
magnetic fields from these vectors we get
$$
\vec B_\uparrow=\vec\nabla\wedge\vec A_\uparrow=-\hat k(\partial_x
W_1+\partial_y W_2)=\hat k, \quad \vec
B_\downarrow=\vec\nabla\wedge\vec A_\downarrow=\hat k(\partial_x
V_1+\partial_y V_2)=-\hat k,
$$
because of the equalities in (\ref{414}). Hence, for any possible
choice of superpotentials, $h'$ and $h$ respectively describe an
electron in an {\em up} and in a {\em down} uniform magnetic
field, as the original hamiltonians $H_0'$ and $H_0$. Incidentally this suggests that we should further analyze this model in
the light of the modular
structure, recently considered in \cite{abh} in the context of SLL.

The following are some possible choices of $\vec W$ and $\vec V$:

\vspace{2mm}

{\bf Choice 1} (SLL). Let us take
$V_1(x,y)=W_1(x,y)=-\frac{x}{2}$, \quad
$V_2(x,y)=W_2(x,y)=-\frac{y}{2}$. If we further take
$\alpha=\gamma=\alpha'=\gamma'=\frac{1}{\sqrt{2}}$ we recover
exactly the usual situation, \cite{alibag}. Moreover, we go back to bosonic rather than pseudo-bosonic commutation relations.

\vspace{2mm}

{\bf Choice 2} (Perturbations of the SLL). First we consider a
symmetric perturbation. For that we take
$V_1(x,y)=-\frac{x}{2}+v_1(y)$, $V_2(x,y)=-\frac{y}{2}+v_2(x)$,
where $v_1$ and $v_2$ are arbitrary (but sufficiently regular)
functions. Hence we get, apart from inessential additive constants,
$W_1(x,y)=-\frac{x}{2}-v_1(y)$, $W_2(x,y)=-\frac{y}{2}-v_2(x)$.
In order not to  trivialize the situation, it is also necessary to
take $v_1(y)$ and $v_2(x)$ complex (at least one of them): this is
the way to get pseudo-bosons rather than {\em simple} bosons.

A non symmetric version of this perturbation can be constructed by
just taking $V_1(x,y)=-a_1\,x+v_1(y)$, $V_2(x,y)=-a_2\,y+v_2(x)$,
with $a_1+a_2=1$.

\vspace{2mm}

{\bf Choice 3} (A general solution). We take
$V_1(x,y)=-x+v_1(y)+\int \frac{\partial V_2(x,y)}{\partial
y}\,dx$, where $V_2(x,y)$ is any function for which this
definition makes sense. In particular, for instance, if we take
$V_2(x,y)=e^{xy}$ then
$V_1(x,y)=-x+v_1(y)+\frac{1}{y^2}\left(x\,y-1\right)e^{xy}$ and,
consequently,
$W_1(x,y)=-v_1(y)-\frac{1}{y^2}\left(x\,y-1\right)e^{xy}$ and
$W_2(x,y)=-y-e^{xy}$.

If we rather take $V_2(x,y)=x^n\,y^k$, $n,k=1,2,3,\ldots$, then
$V_1(x,y)=-x+v_1(y)-\frac{k}{n+1}\,x^{n+1}y^{k-1}$, and so on.

\subsection{A perturbation of the SLL}

We will now focus our attention on Choice 2 above, with an
explicit choice of $v_1(y)$ and $v_2(x)$, and apply the
construction given in Section II. Let
\be
W_1(x,y)=-\frac{x}{2}-ik_1y,\qquad
W_2(x,y)=-\frac{y}{2}-ik_2x,\label{418}
\en
with $k_1$ and $k_2$
real and not both zero (not to go back to
 SLL). In this case the operators in (\ref{415}) assume the following differential expressions: \bea
 \left\{
    \begin{array}{ll}
A'=\alpha'\left(\partial_x-i\partial_y+\frac{x}{2}(1+2k_2)-\frac{iy}{2}(1-2k_1)\right),\\
B'=\gamma'\left(-\partial_x-i\partial_y+\frac{x}{2}(1-2k_2)+\frac{iy}{2}(1+2k_1)\right),\\
A=\alpha\left(-i\partial_x+\partial_y-\frac{ix}{2}(1+2k_2)+\frac{y}{2}(1-2k_1)\right),\\
B=\gamma\left(-i\partial_x-\partial_y+\frac{ix}{2}(1-2k_2)+\frac{y}{2}(1+2k_1)\right).\\
      \end{array}
        \right.\label{419} \ena
In order to check Assumptions 1 and 2 of the previous section, we
first look for vectors $\varphi_{0,0}(x,y)$ and $\Psi_{0,0}(x,y)$
satisfying $A\varphi_{0,0}(x,y)=0$ and
$B^\dagger\Psi_{0,0}(x,y)=0$. We get \bea
 \left\{
    \begin{array}{ll}
\varphi_{0,0}(x,y)=N_\varphi\,\exp\left\{-\frac{x^2}{4}(1+2k_2)-\frac{y^2}{4}(1-2k_1)\right\}\\
\Psi_{0,0}(x,y)=N_\Psi\,\exp\left\{-\frac{x^2}{4}(1-2k_2)-\frac{y^2}{4}(1+2k_1)\right\},\\
\end{array}\right.
\label{420}\ena where $N_\varphi$ and $N_\Psi$ are normalization
constants which are chosen in such a way that
$\left<\varphi_{0,0},\Psi_{0,0}\right>=1$. Of course, in order for
this result to make sense, the two functions must belong to the
Hilbert space $\Hil$ we are considering here, i.e.
$\Lc^2(\Bbb{R}^2)$. This imposes some constraints on $k_1$ and
$k_2$: $-\frac{1}{2}<k_j<\frac{1}{2}$, $j=1,2$.

It is possible to check that the same functions also satisfy
$A'\varphi_{0,0}(x,y)=0$ and $B'^\dagger\Psi_{0,0}(x,y)=0$. It is
now evident that Assumptions 1 and 2 are satisfied. Indeed the
action of, say, $B_1^n$  on $\varphi_{0,0}(x,y)$ simply produces
some polynomial (see (\ref{423}) below) of the $n$-th degree times
a gaussian: this resulting function belongs clearly to
$\Lc^2(\Bbb{R}^2)$ for all $n$. This fact allows us to define the following
functions \be
\varphi_{n,l}(x,y)=\frac{B'^n\,B^l}{\sqrt{n!\,l!}}\,\varphi_{0,0}(x,y),
\quad\mbox{ and }\quad
\Psi_{n,l}(x,y)=\frac{(A'^\dagger)^n\,(A^\dagger)^l}{\sqrt{n!\,l!}}\,\Psi_{0,0}(x,y),
\label{421}\en where $n,l=0,1,2,3,\ldots$. As we have seen in the
previous section, the sets
$\F_\Psi=\{\Psi_{n,l}(x,y),\,n,l\geq0\}$ and
$\F_\varphi=\{\varphi_{n,l}(x,y),\,n,l\geq0\}$ are biorthogonal.
In fact, with our previous choice of the normalization constants,
we have \be
\left<\Psi_{n,l},\varphi_{m,k}\right>=\delta_{n,m}\delta_{l,k},
\quad \forall n,m,l,k\geq0. \label{422}\en Of course these vectors
diagonalize the operators $h=N-\frac{1}{2}\,\1$ and
$h'=N'-\frac{1}{2}\,\1$, as well as their adjoints
$h^\dagger=\N-\frac{1}{2}\,\1$ and
$h'^\dagger=\N'-\frac{1}{2}\,\1$, where $N=BA$, $N'=B'A'$,
$\N=N^\dagger$ and $\N'=N'^\dagger$. We find:
$$
h'\varphi_{n,l}=\left(n-\frac{1}{2}\right)\varphi_{n,l}, \quad
h\,\varphi_{n,l}=\left(l-\frac{1}{2}\right)\varphi_{n,l},
$$
and
$$
h'^\dagger\Psi_{n,l}=\left(n-\frac{1}{2}\right)\Psi_{n,l}, \quad
h^\dagger\Psi_{n,l}=\left(l-\frac{1}{2}\right)\Psi_{n,l}.
$$
The next step consists in proving that the sets $\F_\varphi$ and
$\F_\Psi$ are complete in $\Hil$. This is a consequence of the
fact that (a.) the set
$\F_h:=\{h_{n,m}(x,y):=x^n\,y^m\,\varphi_{0,0}(x,y), \,n,m\geq0\}$
is complete in $\Lc^2(\Bbb{R}^2)$; (b.) each function of $\F_h$,
can be written as a finite linear combination of some
$\varphi_{i,j}(x,y)$. Then it is clear that, if by assumption
$f\in\Hil$ is such that $\left<f,\varphi_{i,j}\right>=0$ for all
$i$ and $j$, then $\left<f,h_{n,m}\right>=0$ for all $n$ and $m$,
so that $f=0$. Of course the same argument allows us to prove that
$\F_\Psi$ is complete in $\Hil$.

This  result implies that also Assumption 3 of Section II is
satisfied. Now we could introduce the intertwining operators
$S_\varphi$ and $S_\Psi$ and check, among other properties, if
they are bounded or not. This is related to the fact that, as we
will first show, the sets $\F_\varphi$ and $\F_\Psi$ are not Riesz
bases, except when $k_1 = k_2 = 0$ (see (\ref{420})). To check this claim, we  introduce the orthonormal basis of
$\Lc^2(\Bbb{R}^2)$ arising from the SLL, \cite{alibag},
$$\F_{\varphi}^{(0)}:=\left\{\varphi_{n,l}^{(0)}(x,y):=\frac{B_0'^n\,B_0^l}{\sqrt{n!\,l!}}
\varphi_{0,0}^{(0)}(x,y),\quad n,m\geq0\right\},$$ where
$\varphi_{0,0}^{(0)}(x,y)=\frac{1}{\sqrt{2\pi}}\,e^{-(x^2+y^2)/4}$
is the vacuum of $A_0=\frac{1}{\sqrt{2}}(Q_0+iP_0)$ and
$A_0'=\frac{1}{\sqrt{2}}(Q_0'+iP_0')$. Recall that, for SLL,
$B'_0=A_0'^\dagger$ and $B_0=A_0^\dagger$.

To prove now that $\F_\varphi$ is not a Riesz basis, we will show
that an operator $T_\varphi$ exists mapping $\F_{\varphi}^{(0)}$
into $\F_{\varphi}$,  that $T_\varphi$ is invertible, but
$T_\varphi$ and/or $T_\varphi^{-1}$ are not bounded. Finding this
operator is simple. Indeed it is easy to first check that \be
\varphi_{n,0}^{(0)}(x,y)=\frac{1}{\sqrt{2^n\,n!}}\,(x+iy)^n\,\varphi_{0,0}^{(0)}(x,y),\quad
\varphi_{0,l}^{(0)}(x,y)=\frac{i^l}{\sqrt{2^l\,l!}}\,(x-iy)^l\,\varphi_{0,0}^{(0)}(x,y)
\label{423}\en and \be
\varphi_{n,0}(x,y)=\frac{\gamma'^n}{\sqrt{n!}}\,(x+iy)^n\,\varphi_{0,0}(x,y),\quad
\varphi_{0,l}(x,y)=\frac{(i\gamma)^l}{\sqrt{l!}}\,(x-iy)^l\,\varphi_{0,0}(x,y),
\label{424}\en for all $n,l\geq0$. Similar formulae are deduced
for $\Psi_{n,0}(x,y)$ and $\Psi_{0,l}(x,y)$. From a comparison
between (\ref{423}) and (\ref{424}) it is clear that $T_\varphi$
can exist only if $\gamma=\gamma'=\frac{1}{\sqrt{2}}$. Assuming
this to be so, we have
\be
\frac{\varphi_{n,0}(x,y)}{\varphi_{n,0}^{(0)}(x,y)}=\frac{\varphi_{0,l}(x,y)}{\varphi_{0,l}^{(0)}(x,y)}=
\frac{\varphi_{0,0}(x,y)}{\varphi_{0,0}^{(0)}(x,y)},
\label{425}\en
for all $n,l\geq0$. This suggest that we define
$T_\varphi$ as the ratio in the right-hand side of this equality:
\be
T_\varphi=\frac{\varphi_{0,0}(x,y)}{\varphi_{0,0}^{(0)}(x,y)}=\sqrt{2\pi}N_\varphi\,
e^{-\frac{x^2}{2}\,k_2+\frac{y^2}{2}\,k_1}. \label{426}\en Of
course we have still to check that with this definition
$\varphi_{n,l}(x,y)=T_\varphi\varphi_{n,l}^{(0)}(x,y)$ holds also
if both $n$ and $l$ are not zero. This can be proven observing
that, for all $n\geq 0$, the following intertwining relation
holds: \be B'^nT_\varphi=T_\varphi(A_0'^\dagger)^n.\label{427}\en
Therefore, since
$$
\varphi_{n,l}(x,y)=T_\varphi\varphi_{n,l}^{(0)}(x,y)\quad
\Leftrightarrow\quad
B'^n\varphi_{0,l}=T_\varphi(A_0'^\dagger)^n\varphi_{0,l}^{(0)}\quad
\Leftrightarrow\quad
B'^nT_\varphi\varphi_{0,l}^{(0)}=T_\varphi(A_0'^\dagger)^n\varphi_{0,l}^{(0)},
$$
our claim immediately follows. Formula (\ref{427}) can be proved
by induction on $n$. The inverse of $T_\varphi$ is
$T_\varphi^{-1}=\frac{1}{\sqrt{2\pi}N_\varphi}\,e^{\frac{x^2}{2}\,k_2-\frac{y^2}{2}\,k_1}$.
It is clear that both $T_\varphi$ and/or $T_\varphi^{-1}$ are
unbounded on $\Lc^2(\Bbb{R}^2)$ for all possible choices of $k_1$ and $k_2$ in $\left(-\,\frac{1}{2},\frac{1}{2}\right)$, except when $k_1 = k_2 = 0$, i.e., in the case of the SLL. Hence, for well known general
reasons, \cite{you,chri}, $\F_\varphi$ cannot be a Riesz basis.

Essentially the same arguments  also show that $\F_\Psi$ is not a
Riesz basis, either. Indeed, an operator $T_\Psi$ mapping
$\F_{\varphi}^{(0)}$ into $\F_{\Psi}$ can be found and it is \be
T_\Psi=\frac{\Psi_{0,0}(x,y)}{\varphi_{0,0}^{(0)}(x,y)}=\sqrt{2\pi}N_\Psi\,e^{\frac{x^2}{2}\,k_2-\frac{y^2}{2}\,k_1}.
\label{428}\en This operator satisfies
$\Psi_{n,l}(x,y)=T_\Psi\varphi_{n,l}^{(0)}(x,y)$ for all possible
choices of $n$ and $l$ greater or equal to zero. Therefore, since
$\varphi_{n,l}(x,y)=T_\varphi\varphi_{n,l}^{(0)}(x,y)= (T_\varphi
T_\Psi^{-1})\Psi_{n,l}(x,y)$, the operators $S_\varphi$ and
$S_\Psi$ in (\ref{213}) can be easily identified and look like \be
S_\varphi=T_\varphi
T_\Psi^{-1}=\frac{N_\varphi}{N_\Psi}\,e^{-x^2k_2+y^2k_1}, \quad
S_\Psi=S_\varphi^{-1}=T_\Psi
T_\varphi^{-1}=\frac{N_\Psi}{N_\varphi}\,e^{x^2k_2-y^2k_1}.
\label{430}\en Notice that for any choice of $k_1$ and $k_2$  in
$\left(-\frac{1}{2},\frac{1}{2}\right)$, other than when $(k_1 , k_2 ) = (0,0)$, at least one of these
operators is unbounded.

\vspace{3mm}

We will now construct a set of {\em bicoherent states}  for our
GLL. However, rather than using the definitions in (\ref{32}), it is
convenient to look for solutions in the $(x,y)-$space  of the
eigenvalue equations \bea
 \left\{
    \begin{array}{ll}
A\tilde\varphi(x,y;z,z')=z\tilde\varphi(x,y;z,z')\\
A'\tilde\varphi(x,y;z,z')=z'\tilde\varphi(x,y;z,z')\\
B^\dagger\tilde\Psi(x,y;z,z')=z\tilde\Psi(x,y;z,z')\\
B'^\dagger\tilde\Psi(x,y;z,z')=z'\tilde\Psi(x,y;z,z')\\
      \end{array}
        \right.\label{431} \ena
where, as suggested by our previous results,  we take
$\alpha=\alpha'=\gamma=\gamma'=\frac{1}{\sqrt{2}}$ in
(\ref{419}).  The square integrable solutions of the differential
equations in (\ref{431}) are \bea
 \left\{
    \begin{array}{ll}
\tilde\varphi(x,y;z,z')=N_A(z,z')\,e^{-[(1+2k_2)x^2-(1-2k_1)y^2]/4}\,e^{\frac{1}{\sqrt{2}}[(z'+iz)x+(z+iz')y]}\\
\tilde\Psi(x,y;z,z')=N_B(z,z')\,e^{-[(1-2k_2)x^2-(1+2k_1)y^2]/4}\,e^{\frac{1}{\sqrt{2}}[(z'+iz)x+(z+iz')y]}\\
      \end{array}
        \right.\; , \label{432} \ena
where $z$ and $z'$ are complex parameters.

The normalization is fixed by requiring that
$$\left<\tilde\varphi(x,y;z,z'),\tilde\Psi(x,y;z,z')\right>_{\Lc^2(\Bbb{R}^2)} =
\left<\varphi(z,z'),\Psi(z,z')\right>_\Hil\; , $$where in the rhs the coherent states introduced in (\ref{32}), and living in the Hilbert space $\Hil$, appear. Notice that $\left<\varphi(z,z'),\Psi(z,z')\right>_\Hil=1$
for all $z$ and $z'$ in $\D$. Then we find, with a suitable
choice of phases,
$$
N_A(z,z')N_B(z,z')=\frac{1}{2\pi}\,e^{-|z-i\overline{z'}|^2}.
$$
Notice that these states reduce to the standard two dimensional
gaussian $\frac{1}{\sqrt{2\pi}}e^{-(x^2+y^2)/4}$ when
$z=z'=k_1=k_2=0$, i.e., for the  SLL and for eigenvalues of the
lowering operators both equal to zero. It is now a straightforward
computation to check the resolution of the identity for these
states \be \frac{1}{\pi^2}\int_{\Bbb{C}^2}dzdz'
|\tilde\varphi(x,y;z,z')><\tilde\Psi(x,y;z,z')|=\1\label{433}\en
where $\1$ is the identity in $\Lc^2(\Bbb{R}^2)$.

This result is by no means in disagreement with the theorem stated in Section II. The first reason is that it is not clear
that the functions $\tilde\varphi(x,y;z,z')$ and
$\tilde\Psi(x,y;z,z')$ coincide with $\varphi(z,z')$ and
$\Psi(z,z')$, for which the theorem was stated. Secondly, and
more important, that theorem gives only a sufficient condition.
Hence, if we would be able to prove that
$\tilde\varphi(x,y;z,z')=\varphi(z,z')$ and
$\tilde\Psi(x,y;z,z')=\Psi(z,z')$,  this computation will
provide a nice counterexample showing that the conditions of the
theorem are, in fact, only sufficient and not necessary. This is work in progress.

\section{Damped harmonic oscillator}

An interesting example of two-dimensional pseudo-bosons is
provided by the damped harmonic oscillator (DHO). In \cite{ban}
the authors have discussed a possible approach to the quantization
of the DHO. This is a non conservative system, so that a
Hamiltonian approach requires a certain amount of care. The approach which
was proposed already in 1977, \cite{fesh}, is to consider
the DHO as a part of a larger system, involving also a second
oscillator which is forced and which takes the energy lost by the
DHO, so that this larger system is conservative. The original
equation of motion, $m\ddot x+\gamma \dot x+kx=0$, is therefore
complemented by a second {\em virtual} equation,  $m\ddot y-\gamma
\dot y+ky=0$, and the classical lagrangian for the system looks
like $L=m\dot x\dot y+\frac{\gamma}{2}(x\dot y-\dot xy)-kxy$,
which corresponds to a classical Hamiltonian $H=p_x\,\dot
x+p_y\,\dot
y-L=\frac{1}{m}\left(p_x+\gamma\frac{y}{2}\right)\left(p_y-\gamma\frac{x}{2}\right)+kxy$,
where $p_x=\frac{\partial L}{\partial\dot x}$ and
$p_y=\frac{\partial L}{\partial\dot y}$ are the conjugate momenta.
The introduction of pseudo-bosons is based on two successive
changes of variables and on a canonical quantization. First of all
we introduce the new variables $x_1$ and $x_2$ via
$x=\frac{1}{\sqrt{2}}(x_1+x_2)$, $y=\frac{1}{\sqrt{2}}(x_1-x_2)$.
Then $L=\frac{1}{2}m\left(\dot x_1^2-\dot
x_2^2\right)+\frac{\gamma}{2}\left(x_2\dot x_1-x_1\dot
x_2\right)-\frac{k}{2}(x_1^2-x_2^2)$ and
$H=\frac{1}{2m}\left(p_1-\gamma\frac{x_2}{2}\right)^2+\frac{1}{2m}\left(p_2+\gamma\frac{x_1}{2}\right)^2+
\frac{k}{2}(x_1^2-x_2^2)$. The second change of variable is the
following:

\bea
 \left\{
    \begin{array}{ll}
p_+=\sqrt{\frac{\omega_+}{2m\Omega}}p_1+i\,\sqrt{\frac{m\Omega\omega_+}{2}}\,x_2,\\
p_-=\sqrt{\frac{\omega_-}{2m\Omega}}p_1-i\,\sqrt{\frac{m\Omega\omega_-}{2}}\,x_2,\\
x_+=\sqrt{\frac{m\Omega}{2\omega_+}}x_1+i\,\sqrt{\frac{1}{2m\Omega\omega_+}}\,p_2,\\
x_-=\sqrt{\frac{m\Omega}{2\omega_-}}x_1-i\,\sqrt{\frac{1}{2m\Omega\omega_-}}\,p_2,\\
      \end{array}
        \right.\label{51} \ena
where we have introduced
$\Omega=\sqrt{\frac{1}{m}\left(k-\frac{\gamma^2}{4m}\right)}$ and
the  two following complex quantities $\omega_\pm=\Omega\pm
i\frac{\gamma}{2m}$. In the rest of the section we will assume
that $k\geq \frac{\gamma^2}{4m}$, so that $\Omega$ is
real. Up to now, we are still at a classical level, so that
$\overline\omega_+=\omega_-$, $\overline p_+=p_-$, $\overline
x_+=x_-$, and consequently, see below, $\overline H_+=H_-$ and $\overline H=H$. Hence
$H$ is a real Hamiltonian. Indeed, with these definitions, the Hamiltonian
looks like the hamiltonian of a two-dimensional harmonic
oscillator
$$
H=\frac{1}{2}\left(p_+^2+\omega_+^2x_+^2\right)+\frac{1}{2}\left(p_-^2+\omega_-^2x_-^2\right)=:H_++H_-
$$
at least formally.

At this stage we quantize canonically  the system,
\cite{ban}: we require that the following commutators are
satisfied: \be [x_+,p_+]=[x_-,p_-]=i\1, \label{52}\en all the other
commutators being trivial. We also have to require that $
p_+^\dagger=p_-$ and that $x_+^\dagger=x_-$, which are the quantum
version of the {\em compatibility} conditions above.  The
pseudo-bosons now appear:
 \bea
 \left\{
    \begin{array}{ll}
a_+=\sqrt{\frac{\omega_+}{2}}\left(x_++i\,\frac{p_+}{\omega_+}\right),\\
a_-=\sqrt{\frac{\omega_-}{2}}\left(x_-+i\,\frac{p_-}{\omega_-}\right),\\
b_+=\sqrt{\frac{\omega_+}{2}}\left(x_+-i\,\frac{p_+}{\omega_+}\right),\\
b_-=\sqrt{\frac{\omega_-}{2}}\left(x_--i\,\frac{p_-}{\omega_-}\right),\\
      \end{array}
        \right.\label{53} \ena
and indeed we have $[a_+,b_+]=[a_-,b_-]=\1$, all the other
commutators being zero. Notice also that $b_+=a_-^\dagger$ and
$b_-=a_+^\dagger$. Moreover $H$ can be written in term of the
operators $N_\pm=b_\pm a_\pm$ as
$H=\omega_+N_++\omega_-N_-+\frac{\omega_++\omega_-}{2}\,\1$. So the
hamiltonian of the quantum DHO is simply
written in terms of pseudo-bosonic operators.

\subsection{About Assumptions 1-3}

This system provides a non trivial example of pseudo-bosonic
operators which do not satisfy any of the Assumptions 1-3 of
Section II. To show this, we first observe that a possible
representation of the operators in (\ref{52}) is the following \bea
 \left\{
    \begin{array}{ll}
x_+=\frac{1}{\Gamma\,\overline{\delta}-\delta\,\overline{\Gamma}}\left(\overline{\Gamma}\,p_y+\overline{\delta}\,x\right),\\
x_-=\frac{-1}{\Gamma\,\overline{\delta}-\delta\,\overline{\Gamma}}\left({\Gamma}\,p_y+{\delta}\,x\right),\\
p_+=\Gamma \,p_x+\delta \,y,\\
p_-=\overline{\Gamma} \,p_x+\overline{\delta} \,y,\\
      \end{array}
        \right.\label{54} \ena
for all choices of $\Gamma$ and $\delta$ such that
$\Gamma\,\overline{\delta}\neq \delta\,\overline{\Gamma}$. Here
$x$, $y$, $p_x$ and $p_y$ are pairwise conjugate self-adjoint
operators: $[x,p_x]=[y,p_y]=i\1$. Notice that these operators also
satisfy the compatibility conditions $ p_+^\dagger=p_-$ and
$x_+^\dagger=x_-$. Hence it is natural to represent $x$ and $y$ as
the standard multiplication operators and $p_x$ and $p_y$ as
$-i\,\frac{\partial}{\partial\,x}=-i\,\partial_x$ and
$-i\,\frac{\partial}{\partial\,y}=-i\,\partial_y$. Then we get

\bea
 \left\{
    \begin{array}{ll}
a_+=\sqrt{\frac{\omega_+}{2}}\,\left\{\left(\beta\,x+i\,\frac{\delta}{\omega_+}\,y\right)+\left(\frac{\Gamma}{\omega_+}\,
\partial_x-i\,\alpha\,\partial_y\right)\right\},\\
a_-=\sqrt{\frac{\omega_-}{2}}\,\left\{\left(\overline{\beta}\,x+i\,\frac{\overline{\delta}}{\omega_-}\,y\right)+
\left(\frac{\overline{\Gamma}}{\omega_-}\,\partial_x-i\,\overline{\alpha}\,\partial_y\right)\right\},\\
b_+=\sqrt{\frac{\omega_+}{2}}\,\left\{\left(\beta\,x-i\,\frac{\delta}{\omega_+}\,y\right)-\left(\frac{\Gamma}{\omega_+}\,
\partial_x+i\,\alpha\,\partial_y\right)\right\},\\
b_-=\sqrt{\frac{\omega_-}{2}}\,\left\{\left(\overline{\beta}\,x-i\,\frac{\overline{\delta}}{\omega_-}\,y\right)-
\left(\frac{\overline{\Gamma}}{\omega_-}\,\partial_x+i\,\overline{\alpha}\,\partial_y\right)\right\},\\
      \end{array}
        \right.\label{55} \ena
where, to simplify the notation, we have introduced
$\alpha=\frac{\overline{\Gamma}}{\Gamma\,\overline{\delta}-\delta\,\overline{\Gamma}}$
and
$\beta=\frac{\overline{\delta}}{\Gamma\,\overline{\delta}-\delta\,\overline{\Gamma}}$.

\vspace{2mm}

{\bf Remark:--} a different representation of $x_\pm$ and $p_\pm$ could be deduced using the results of Section III. However,  while the pseudo-bosonic commutation rules would be easily recovered, the compatibility conditions $x_+^\dagger=x_-$ and $p_+^\dagger=p_-$ would be lost. Hence this choice is not compatible with our requirements.

\vspace{2mm}

Assumption 1 of Section II requires the existence of a
square-integrable function $\varphi_{0,0}(x,y)$ such that, first
of all, $a_+\varphi_{0,0}(x,y)=a_-\varphi_{0,0}(x,y)=0$.
Analogously, Assumption 2  requires the existence of a (possibly
different) square-integrable function $\Psi_{0,0}(x,y)$ such that,
first of all,
$b_+^\dagger\Psi_{0,0}(x,y)=b_-^\dagger\Psi_{0,0}(x,y)=0$.
However, since $b_+=a_-^\dagger$ and $b_-=a_+^\dagger$, these two
functions, if they exist, satisfy the same differential equations.
Hence, apart from  a normalization constant, we can chose them to be
coincident.  It is possible to check that a solution of
$a_+\varphi_{0,0}(x,y)=a_-\varphi_{0,0}(x,y)=0$ is the following:
\be
\varphi_{0,0}(x,y)=N_0\,\exp\left\{-\,\frac{\beta\,\omega_+}{2\,\Gamma}\,x^2+\frac{\delta}{2\,\alpha\,\omega_+}\,y^2\right\}.
\label{56}\en Notice that, in order for this function to be a solution
of both $a_+\varphi_{0,0}(x,y)=0$ and $a_-\varphi_{0,0}(x,y)=0$ it
is necessary and sufficient to have the following identity
satisfied:
$\frac{\omega_+}{\omega_-}=-\,\frac{\delta}{\overline{\delta}}\,\frac{\Gamma}{\overline{\Gamma}}$.
This is not a big requirement, clearly. What is crucial, on the
other hand, is that the function $\varphi_{0,0}(x,y)$, and
$\Psi_{0,0}(x,y)$ should consequently be square integrable.
This is possible only if
$\Re\left(\frac{\beta\,\omega_+}{2\Gamma}\right)>0$ and if, at the
same time, $\Re\left(\frac{\delta}{\alpha\,\omega_+}\right)<0$.
Now, it is not hard to check that these two conditions are
incompatible: if one is verified, the other is not. Therefore the
conclusion is that, following the procedure we have considered so
far, Assumptions 1 and 2 are violated and, of course, Assumption 3
cannot even be considered since it is meaningless. Of course
this does not mean that for the quantum DHO
the construction proposed in Section II cannot be considered. It
only means that with the choices we have considered here, this is
not possible. It could be possible, however, to look for some
different representation of the operators, satisfying the compatibility condition, and see
if it is possible to satisfy Assumptions 1, 2
and 3. This is work in progress.

\section{Conclusions}

In this paper we have constructed a physically motivated
two-dimensional family of pseudo-bosons arising from a generalized
version of the Landau levels. This generalization has been shown
to be essentially a gauge transformation. Coherent states have
been constructed and the resolution of the identity has been
proved.

We  have also considered a quantum damped harmonic oscillator: this
provides a nice example of a pseudo-bosonic system for which all
the assumptions of Section II are violated. In conclusion, many examples exist, see Section III and references \cite{bagpb1,bagpb2,tri,baglast} among the others, in which Assumptions 1-3, and sometimes Assumption 4, are satisfied. But other examples exist as well for which, even if pseudo-bosonic commutation rules are recovered, none of the Assumptions hold true. This suggests to take care explicitly of these Assumptions when dealing with pseudo-bosons.

\section*{Acknowledgements}
   The authors would like to acknowledge financial support from the
   Universit\`a di Palermo through Bando CORI, cap. B.U. 9.3.0002.0001.0001.
One of us (STA) would like to acknowledge a grant from the Natural
Sciences and Engineering Research Council (NSERC) of Canada.

\end{document}